\documentstyle[aps,prl,preprint,tighten]{revtex}

\def\dalemb#1#2{{\vbox{\hrule height .#2pt
        \hbox{\vrule width.#2pt height#1pt \kern#1pt
                \vrule width.#2pt}
        \hrule height.#2pt}}}
\def\square{\mathord{\dalemb{6.8}{7}\hbox{\hskip1pt}}}

\let\la=\label
\newcommand{\be}{\begin{equation}}
\newcommand{\ee}{\end{equation}}
\newcommand{\bea}{\begin{eqnarray}}
\newcommand{\eea}{\end{eqnarray}}

\begin{document}
\preprint{\vtop{\hbox{UM-TH-00-12}\hbox{hep-th/0003237}\vskip24pt}}

\title{Complementarity of the Maldacena and Randall-Sundrum
Pictures\footnote{Research supported in part by NSF Grant PHY-9722090.}}

\author{M.~J.~Duff\footnote{mduff@umich.edu} and
James T.~Liu\footnote{jimliu@umich.edu}}

\address{Randall Laboratory, Department of Physics, University of Michigan,\\
Ann Arbor, MI 48109--1120}

\maketitle

\begin{abstract}
We revive an old result, that one-loop corrections to the graviton propagator
induce $1/r^{3}$ corrections
to the Newtonian gravitational potential, and compute the coefficient due to
closed loops of the U(N) ${\cal N}=4 $ super-Yang-Mills theory that 
arises in Maldacena's AdS/CFT correspondence. We find exact agreement
with the coefficient appearing in the Randall-Sundrum brane-world proposal.
This provides more evidence for the complementarity of the two pictures.
\end{abstract}


\narrowtext

%%%%%%%%%%%%%%%%%%%%%%%%%%%%%%%%%%%%%%%%%%%

It is an old, and seemingly forgotten,
result that one-loop corrections to the graviton propagator induce
$1/r^{3}$ corrections to the gravitational potential \cite{Duff1,Duff2}:
\be
V(r)= \frac{Gm_{1}m_{2}}{r}\biggl
(1+\frac{\alpha G}{r^{2}} \biggr),
\la{Newton}
\ee
where $G$ is the four-dimensional Newton's constant, ${\hbar}=c=1$ and
$\alpha$ is a purely numerical coefficient given, in the case of spins
$s\leq1$, by $45 \pi \alpha=12N_{1}+3N_{1/2}+N_{0}$, where $N_{s}$ are
the numbers of particle species of spin $s$ going around the loop
\cite{Capper1,Capper2,Capper4}. However, the importance of
this result has recently become apparent in attempts
\cite{Verlinde,Witten,Gubser,Giddings,Hawking} to relate two topical
but, at first sight,
different developments in quantum gravity. These are Maldacena's AdS/CFT
correspondence \cite{Maldacena,Wittenads,Gubserklebanovpolyakov}
and the Randall-Sundrum brane-world mechanism \cite{Randall}.

The AdS/CFT correspondence in general relates the gravitational dynamics of a
$(d+1)$-dimensional anti-de Sitter spacetime, AdS$_{d+1}$, to a $d$-dimensional
conformal field theory, CFT$_d$.  In the case of $d=4$, Maldacena's
conjecture, based on the decoupling limit of D3-branes in Type $IIB$ string
theory compactified on $S^{5}$, then relates the dynamics of AdS$_5$ to
an ${\cal N}=4$ superconformal $U(N)$ Yang-Mills theory on its
four-dimensional boundary \cite{Maldacena}. Other compactifications are 
also possible, leading to different SCFT's on the boundary. We note that,
by choosing Poincar\'e coordinates on AdS$_5$, the metric may be written as
\begin{equation}
ds^2=e^{-2y/L}(dx^\mu)^2+dy^2,
\end{equation}
in which case the superconformal Yang-Mills theory is taken to reside at the
boundary $y\to-\infty$.

The Randall-Sundrum mechanism, on the other hand, was originally
motivated, not via the decoupling of gravity from D3-branes,
but rather as a possible mechanism for evading Kaluza-Klein compactification
by localizing gravity in the presence of an uncompactified extra dimension.
This was accomplished by inserting a positive tension 3-brane (representing
our spacetime) into AdS$_5$.  In terms of the Poincar\'e patch of AdS$_5$
given above, this corresponds to removing the region $y<0$, and either
joining on a second partial copy of AdS$_5$, or leaving the brane at the end
of a single patch of AdS$_5$.  In either case
the resulting Randall-Sundrum metric is given by
\begin{equation}
ds^2=e^{-2|y|/L}(dx^\mu)^2+dy^2,
\end{equation}
where $y\in(-\infty,\infty)$ or $y\in[0,\infty)$ for a `two-sided' or
`one-sided' Randall-Sundrum brane respectively.

The similarity of these two scenarios led to the notion that they
are in fact closely tied together.  To make this connection clear,
consider the one-sided Randall-Sundrum brane.  By introducing a boundary
in AdS$_5$ at $y=0$, this model is conjectured to be dual to a cutoff CFT
coupled to gravity, with $y=0$, the location of the Randall-Sundrum
brane, providing the UV cutoff.  This extended version of the Maldacena
conjecture \cite{wittensusskind}
then reduces to the standard AdS/CFT duality as the boundary
is pushed off to $y\to-\infty$, whereupon the cutoff is removed and gravity
becomes completely decoupled.  Note in particular that this connection
involves a single CFT at the boundary of a single patch of AdS$_5$.  For the
case of a brane sitting between two patches of AdS$_5$, one would instead
require two copies of the CFT, one for each of the patches.

It has been suggested \cite{Witten,Gubser,Giddings} that a crucial test of
this Randall-Sundrum version of the Maldacena conjecture would be to compare
the $1/r^{3}$ corrections to Newton's law in both pictures. From the above,
we see that the contribution of a single CFT, with
$(N_{1},N_{1/2},N_{0})=(N^{2},4N^{2},6N^{2})$, is
\be
V(r)= \frac{Gm_{1}m_{2}}{r}\biggl
(1+\frac{2N^{2}G}{3\pi r^{2}} \biggr).
\la{AdS}
\ee
Using the AdS/CFT relation $N^{2}=\pi L^{3}/2G_{5}$ \cite{Maldacena} and the
one-sided brane-world relation $G=2G_{5}/L$ \cite{Randall,Gubser},
where $G_{5}$ is the five-dimensional
Newton's constant and $L$ is the radius of AdS$_{5}$, this becomes
\be
V(r)= \frac{Gm_{1}m_{2}}{r}\biggl(1+\frac{2L^{2}}{3 r^{2}} \biggr).
\la{RS}
\ee
The coefficient of the $1/r^{3}$ term is $2/3$ of the Randall-Sundrum result
quoted in \cite{Randall}, but in fact agrees with the more thorough analysis
of \cite{Garriga}.  We shall confirm below that a more careful
analysis of the Randall-Sundrum picture using the results of
\cite{Giddings,Muck} yields exactly the same answer as the above AdS/CFT
calculation, thus providing strong evidence for the conjectured duality
of the two pictures.

First we derive (\ref{AdS}) in more detail by computing the lowest
order quantum corrections to solutions of Einstein's equations.  Working
with linearized gravity, we begin by writing the metric as
\begin{equation}
g_{\mu\nu}=\eta_{\mu\nu}+h_{\mu\nu},
\end{equation}
so that
\be
\sqrt{-g}g^{\mu\nu} \equiv {\tilde g}^{\mu\nu}
=\eta^{\mu\nu}-\tilde{h}^{\mu\nu}+\cdots,
\ee
where
\begin{equation}
\tilde{h}_{\mu\nu}=h_{\mu\nu}
-{\textstyle{1\over2}}\eta_{\mu\nu}h^\alpha{}_\alpha.
\end{equation}
In harmonic gauge, $\partial_\mu{\tilde g}^{\mu\nu}=0$ ({\it i.e.}
$\partial_\mu\tilde{h}^{\,\mu\nu}=0$), the classical
linearized Einstein equation reads
\begin{equation}
\square\,\tilde{h}^{\,c}_{\mu\nu}(x)=-16\pi G T_{\mu\nu}^{\vphantom{c}}(x),
\end{equation}
where the superscript $c$ denotes the classical contribution.
Fourier transforming to momentum space results in
\begin{equation}
\label{eq:lineins}
\tilde{h}^{\,c}_{\mu\nu}(p)=-16\pi G\Delta_4(p)T_{\mu\nu}^{\vphantom{c}}(p),
\end{equation}
where $\Delta_4(p)=-1/p^2$ is the four-dimensional massless scalar
propagator.

Incorporating one-loop corrections, the quantum corrected metric becomes
\begin{equation}
\tilde{h}_{\mu\nu}^{\vphantom{c}}
=\tilde{h}^{\,c}_{\mu\nu}+\tilde{h}^{\,q}_{\mu\nu},
\end{equation}
where the quantum correction $\tilde{h}_{q}^{\mu\nu}$ is given in
momentum space by
\be
\tilde{h}_{q}^{\,\mu\nu}(p)=D^{\mu\nu\alpha\beta}(p)
\Pi_{\alpha\beta\gamma\delta}(p) \tilde{h}_{c}^{\,\gamma\delta}(p).
\label{eq:qc}
\ee
$D^{\mu\nu\alpha\beta}$ is the graviton propagator,
\be
D^{\mu\nu\alpha\beta}(p)=\frac{1}{2}\Delta_4(p)
        (\eta^{\mu\alpha}\eta^{\nu\beta}+ \eta^{\mu\beta}\eta^{\nu\alpha}
                -\eta^{\mu\nu}\eta^{\alpha\beta}+\cdots),
\la{propagator}
\ee
and $\Pi_{\alpha\beta\gamma\delta}$ is the one-loop graviton self-energy,
which by symmetry and Lorentz invariance must be of the general form
\widetext
\begin{eqnarray}
\Pi_{\alpha\beta\gamma\delta}(p)&=&
p^4\Bigl[\Pi_{1}(p^{2})\eta_{\alpha\beta}\eta_{\gamma\delta}+
\Pi_{2}(p^{2})(\eta_{\alpha\gamma}\eta_{\beta\delta}+
                       \eta_{\alpha\delta}\eta_{\beta\gamma})+
\Pi_{3}(p^{2})(\eta_{\alpha\beta}\hat p_{\gamma}\hat p_{\delta}+
                      \eta_{\gamma\delta}\hat p_{\alpha}\hat p_{\beta})
\nonumber\\
&&\kern6pt+\Pi_{4}(p^{2})(\eta_{\alpha\gamma}\hat p_{\beta}\hat p_{\delta}+
                       \eta_{\alpha\delta}\hat p_{\beta}\hat p_{\gamma}+
                       \eta_{\beta\gamma}\hat p_{\alpha}\hat p_{\delta}+
                       \eta_{\beta\delta}\hat p_{\alpha}\hat p_{\gamma})+
\Pi_{5}(p^{2})\hat p_{\alpha}\hat p_{\beta}\hat p_{\gamma}\hat p_{\delta}
\Bigr].
\label{eq:se}
\end{eqnarray}
The ellipses in (\ref{propagator}) refer to gauge dependent terms in the
propagator which make no contribution if coupled to conserved sources.
Combining (\ref{eq:qc}), (\ref{propagator}) and (\ref{eq:se}),
one thus obtains the quantum corrected metric in the form
\begin{equation}
h^q_{\mu\nu}(p)=
-p^2\Bigl[2\Pi_{2}(p){\delta^\alpha_{\mu}}{\delta^{\beta}_{\nu}}
+\Pi_{1}(p)\eta_{\mu \nu}^{\vphantom{\alpha}}
\eta^{\alpha\beta}_{\vphantom{\mu}}
+(\Pi_{3}(p)+\cdots)\hat p_\mu^{\vphantom{\alpha}}
\hat p_\nu^{\vphantom{\alpha}}\eta^{\alpha \beta}_{\vphantom{\mu}} \Bigr]
\tilde{h}^{\,c}_{\alpha \beta},
\la{chi}
\end{equation}
\narrowtext
where non-physical gauge-dependent terms have again been dropped.
Finally, combining both classical and one-loop quantum results at the
linearized level yields
\begin{eqnarray}
\label{eq:pcft}
h_{\mu\nu}(p)&=&-16\pi G\Delta_4(p)
[T_{\mu\nu}(p)-{\textstyle{1\over2}} \eta_{\mu\nu}T^\alpha{}_\alpha(p)]\\
&&-16\pi G[2\Pi_2(p)T_{\mu\nu}(p)+\Pi_1(p)\eta_{\mu\nu}T^\alpha{}_\alpha(p)].
\nonumber
\end{eqnarray}
Note that we have ignored the gauge-dependent term in $h_{\mu\nu}$
proportional to $\hat p_\mu\hat p_\nu$. It makes no contribution when
$h_{\mu\nu}$ is attached to a conserved source $T_{\mu\nu}$ satisfying
$p^\mu T_{\mu \nu} = p^\nu T_{\mu \nu} = 0$.

The actual form of the one-loop $\Pi_i$'s depend on the theory at hand.
However for any massless theory in four-dimensions, after cancelling the
infinities with the appropriate counterterms, the finite remainder must
necessarily have the form
\be
\Pi_{i}(p)=32\pi G
\biggl(a_{i}\;\ln\frac{p^{2}}{\mu^{2}} +b_{i} \biggr),
\la{selfenergy}
\ee
where $a_{i}$ and $b_{i}$, $(i=1,2,3,4,5)$, are numerical coefficients
and $\mu$ is an arbitrary subtraction constant having the dimensions
of mass.  In order to make connection with the Newtonian potential, we
Fourier transform (\ref{eq:pcft}) back to coordinate space.  For the
static potential we obtain the expected $1/r$ behavior at the classical
level, while the quantum term generates the claimed $1/r^3$ correction.
In addition, the
constant parts in (\ref{selfenergy}) give rise to a regulator-dependent
$\delta^3({\bf r})$ contact interaction.  However we have no real
expectation that this one-loop perturbative result remains valid when
continued down to zero size.  Moreover, possible $r^{-3} \ln \mu r$
terms come only from the $\hat p_\mu \hat p_\nu$ terms in (\ref{chi})
and hence drop out. For a point source,
$T_{00}(x)=m\delta^3({\bf r})$, we obtain to this order
\begin{eqnarray}
g_{00}&=&-\biggl(1-\frac{2Gm}{r}-\frac{2\alpha G^{2}m}{r^{3}}\biggr),
\nonumber\\
g_{ij}&=&\biggl(1 + \frac{2Gm}{r}+ \frac{2\beta G^{2}m}{r^{3}}\biggr)
\delta_{ij},
\la{hAdS}
\end{eqnarray}
where, in agreement with \cite{Duff2}, $\alpha=4\cdot32\pi(a_{1}+2a_{2})$ and 
$\beta=-4\cdot32\pi a_{1}$.
This yields the potential given in (\ref{Newton}). Explicit calculations of
the self-energy (\ref{selfenergy}) for spin 1 \cite{Capper1},
spin 1/2 (two-component fermions) \cite{Capper2} and (real conformally
coupled) spin 0 \cite{Capper4} yield%
\footnote{Note that a symmetry factor of $1/2$ was omitted in
Ref.~\cite{Capper1}; this was subsequently corrected in Ref.~\cite{Capper4}}
\begin{eqnarray}
\label{eq:coefs}
a_{i}(s=1)=4a_{i}(s=1/2)&&=12a_{i}(s=0)\nonumber\\
&&\kern-1em =\frac{1}{120(4\pi)^{2}}(-2,3,2,-3,4).
\end{eqnarray}
Note that all spins contribute with the same sign as they must by general
positivity arguments on the self-energy \cite{Capper2}. Thus
\be
\alpha=2\beta=\frac{1}{45\pi}(12N_{1}+3N_{1/2}+N_{0})=\frac{2N^{2}}{3\pi},
\ee
as quoted in the introductory paragraph above.

This $\alpha$ coefficient also determines that part of the Weyl anomaly
\cite{Capper3,Deser} involving the square of the
Weyl tensor \cite{Duff3,Duff4}:
\be
g_{\mu\nu}\langle T^{\mu\nu}\rangle
= b\left (F+\frac{2}{3} \,{\square\, R} \right) +b'G,
\ee
where
\begin{eqnarray}
F&=&C_{\mu\nu\rho\sigma}C^{\mu\nu\rho\sigma}
=R_{\mu\nu\rho\sigma}R^{\mu\nu\rho\sigma}-2R_{\mu\nu}R^{\mu\nu}+
R^{2},\\
G&=&*R_{\mu\nu\rho\sigma}*\!R^{\mu\nu\rho\sigma}
=R_{\mu\nu\rho\sigma}R^{\mu\nu\rho\sigma}-4R_{\mu\nu}R^{\mu\nu}+
\frac{1}{3}R^{2},\nonumber
\end{eqnarray}
and where $b$ and $b'$ are constants
\begin{eqnarray}
b&=&\frac{1}{120(4\pi)^{2}}[12N_{1}+3N_{1/2}+N_{0}],\nonumber\\
b'&=&-\frac{1}{720(4\pi)^{2}}[124N_{1}+11N_{1/2}+2N_{0}].
\end{eqnarray}
Note that for the ${\cal N}=4$ SCFT, the
coefficient of the (Riemann)$^{2}$ term, $b+b'$, vanishes \cite{Duff4}.
The same result is obtained if one calculates the holographic Weyl anomaly
using the AdS/CFT correspondence \cite{Henningson}.
Thus $b=3\alpha/128\pi=c/(4\pi)^{2}$, where the $c$ is the central charge
given in the normalization of \cite{Gubser}. For the central charge,
one obtains $c=\pi L^{3}/8G_{5}$ \cite{Henningson}, so that
\be
G\alpha=\frac{GL^{3}}{3G_{5}}=\frac{2L^{2}}{3},
\la{result}
\ee
where the second equality makes use of the brane-world relation
$G=2G_5/L$.  Although we have focused on the ${\cal N}=4$ SCFT to relate
the coefficient appearing in Newton's law to the central charge, 
the result (\ref{result}) is universal, being independent of which particular
CFT appears in the AdS/CFT correspondence, which is just as well since the
Randall-Sundrum coefficient does not depend on the details of the fields
propagating on the brane.

We now turn to this brane-world, where the five-dimensional action has
the form \cite{Randall}
\begin{equation}
S=\int d^5x\sqrt{-g_{(5)}}[M^3R_{(5)}-\Lambda]+\int
d^4x\sqrt{-g_{(4)}}{\cal L}_{\rm brane}.
\end{equation}
Here $M$ is the five-dimensional Planck mass, $M^3=1/(16\pi G_5)$, and
$\Lambda$ is the cosmological constant in the bulk.
Small fluctuations of the metric
on the brane may be represented by \cite{Randall,Giddings}
\begin{equation}
ds^2=e^{-2|y|/L}[\eta_{\mu\nu}+h_{\mu\nu}(x,y)]dx^\mu dx^\nu+dy^2,
\end{equation}
where $L$ is the `radius' of AdS,
\begin{equation}
R^{(5)}_{MNPQ}=-{1\over L^2}(g^{(5)}_{MP}g^{(5)}_{NQ}
-g^{(5)}_{MQ}g^{(5)}_{NP}),
\end{equation}
and is related to $\Lambda$ by $\Lambda=-12M^3/L^2$.
The brane-world geometry has been chosen such that $x^\mu$ are
coordinates along the 3-brane, while $y$ is the
coordinate perpendicular to the brane (which sits at $y=0$).

Both brane and bulk quantities are contained in the linearized metric
$h_{\mu\nu}(x,y)$.  However, for comparison with the CFT on the brane,
we are only concerned with the former.  Hence we consider a matter source
on the brane, and examine $h_{\mu\nu}(x)\equiv h_{\mu\nu}(x,y=0)$.
For this case, the results of \cite{Giddings,Muck} indicate
\begin{eqnarray}
\label{eq:linrs}
h_{\mu\nu}(p)&=&-{2\over LM^3}\Delta_4(p)
[T_{\mu\nu}(p)-{\textstyle{1\over2}}\eta_{\mu\nu}
T^\alpha{}_\alpha(p)]\nonumber\\
&&-{1\over M^3}\Delta_{KK}(p)[T_{\mu\nu}(p)
-{\textstyle{1\over3}}\eta_{\mu\nu}T^\alpha{}_\alpha(p)].
\end{eqnarray}
This expression has a clear physical meaning; $\Delta_4(p)$, the
four-dimensional massless propagator, corresponds to the zero-mode
graviton localized on the brane, while
\begin{equation}
\Delta_{KK}(p) = -{1\over p}{K_0(pL)\over K_1(pL)}
\end{equation}
is the propagator for the continuum Kaluza-Klein graviton modes.  Comparing
the first term of (\ref{eq:linrs}) to (\ref{eq:pcft}), we obtain the
relation between four- and five-dimensional Newton's constants,
$G=2G_5/L=1/(8\pi LM^3)$ given above.  Note that in the above we have
taken the brane to be at the end of a single patch of AdS$_5$, as was
done in \cite{Gubser,Giddings}.  This corresponds to the case at hand,
since the AdS/CFT relations we have employed above pertain to a single
copy of AdS$_5$.

The continuum graviton modes give rise to corrections to the Newtonian
potential.  At large distances, corresponding to $pL\ll1$, a small
argument expansion for Bessel functions yields
\begin{equation}
\Delta_{KK}(p)={L\over2}\biggl(\ln{p^2L^2\over4}+2\gamma\biggr)
+{\cal O}(p^2),
\end{equation}
and, just as in (\ref{selfenergy}), is the source of the $1/r^3$ correction
to the Newtonian potential.
For a static gravitational source of mass $m$ on the brane,
$T_{00}(p)=2\pi\delta(p_0)m$, evaluating the
Fourier transform for $r\gg L$ yields the linearized metric
\cite{Garriga}
\begin{eqnarray}
h_{00}&=&{2Gm\over r}\biggl(1+{2L^2\over3r^2}+\cdots\biggr),\nonumber\\
h_{ij}&=&{2Gm\over r}\biggl(1+{L^2\over3r^2}+\cdots\biggr)\delta_{ij},
\la{hRS}
\end{eqnarray}
from which one may read off the Newtonian potential (\ref{RS}).

Moreover, all the metric components in (\ref{hRS}) agree with those of
(\ref{hAdS}) and not merely the $g_{00}$ component.  In momentum space, this
may be traced to the behavior of $h_{\mu\nu}$ in the two pictures, namely
(\ref{eq:pcft}) and (\ref{eq:linrs}).  In (\ref{eq:linrs}) the factor 
of $-1/3$ in the non-leading term, as compared with factor $-1/2$ in 
the leading term, is attributable to the fact that the 
Kaluza-Klein gravitons are massive. Whereas in (\ref{eq:pcft}), it is 
because the CFT requires loop corrections with
$\Pi_2(p)=-{3\over2}\Pi_1(p)$, which is in fact satisfied, as far as the
$\ln p^2$ term is concerned, since $a_2=-{3\over2}a_1$. 

We have thus demonstrated that the $1/r^3$ corrections to Newton's law are
identical between the Maldacena and Randall-Sundrum pictures.  This was
examined in the context of a single CFT corresponding to a one-sided
brane-world scenario.  Had we chosen instead to take the brane-world to be 
sitting between two patches AdS$_5$ (one on either side), as was the case
considered in \cite{Randall,Hawking}, we would have obtained a
factor of two in the relation between Newton's constants, with a
corresponding factor in the propagator, (\ref{eq:linrs}).  While
this would ensure the correct four-dimensional behavior of gravity,
given in (\ref{hRS}), the two-sided brane-world relation $G=G_5/L$
will modify the comparison with the one-loop CFT result, (\ref{result}).  To
compensate for this mismatch, one may assume that the two-sided brane-world
is dual to two copies of the CFT coupled to gravity, as is implicit in
\cite{Hawking}.  This leads to the natural picture that a one-sided brane
corresponds to a single CFT while a two-sided brane corresponds to two CFTs.

An intriguing feature of this comparison of the gravitational potential in
both pictures is a highlighting of the classical/quantum nature of
this duality, as seen in the relation
\begin{equation}
\Pi_2(p) + {\cal O}(G^2) = {L\over4}\Delta_{KK}(p).
\end{equation}
The propagator for the continuum graviton modes in the Randall-Sundrum
picture thus incorporates all quantum effects of matter on the brane.
It may be worthwhile to examine this relation at the two-loop or higher
level.  Nevertheless, this agreement at one-loop lends strong support to
the conjectured duality between the two pictures.

\bigskip

We would like to acknowledge the referees for providing valuable
comments.
JTL wishes to thank I.~Giannakis and H.C.~Ren for fruitful discussions
on linearized gravity in the brane-world.

%%%%%%%%%%%%%%%%%%%%%%%%%%%%%%%%%%%%%%%%%%%


\begin{references}

\bibitem{Duff1}
M. J. Duff,
{\it Problems in the classical and quantum theories of gravitation},
Ph. D. thesis, Imperial College, London (1972).

\bibitem{Duff2}
M. J. Duff,
{\sl Quantum corrections to the Schwarzschild solution},
Phys. Rev. {\bf D9}, 1837 (1974).

\bibitem{Capper1}
D. M. Capper, M. J. Duff and L. Halpern,
{\sl Photon corrections to the graviton propagator},
Phys. Rev. {\bf D10}, 461 (1974).

\bibitem{Capper2}
D. M. Capper and M. J. Duff,
{\sl One-loop neutrino corrections to the graviton propagator},
Nucl. Phys. {\bf B82}, 147 (1974).

\bibitem{Capper4}
D. M. Capper,
{\sl On quantum corrections to the graviton propagator},
Nuovo Cimento {\bf A25}, 29 (1975).

\bibitem{Verlinde}
H. Verlinde,
{\sl Holography and compactification},
hep-th/9906182.

\bibitem{Witten}
The complementarity and the Newton's law check were suggested independently
by E. Witten and J. Maldacena in unpublished work.

\bibitem{Gubser}
S. S. Gubser,
{\sl AdS/CFT and gravity},
hep-th/9912001.

\bibitem{Giddings}
S. B. Giddings, E. Katz and L. Randall,
{\sl Linearized gravity in brane backgrounds},
hep-th/0002091.

\bibitem{Hawking}
S. W. Hawking, T. Hertog and H. S. Reall,
{\sl Brane new world},
hep-th/0003052.

\bibitem{Maldacena}
J. Maldacena,
{\sl The large $N$ limit of superconformal field theories and supergravity},
Adv. Theor. Math. Phys. {\bf 2}, 231 (1998), hep-th/9711200.

\bibitem{Wittenads}
E. Witten,
{\sl Anti-de Sitter space and holography},
Adv. Theor. Math. Phys. {\bf 2}, 253 (1998), hep-th/9802150.

\bibitem{Gubserklebanovpolyakov}
S. S. Gubser, I. R. Klebanov and A. M. Polyakov,
{\sl Gauge theory correlators from non-critical string theory},
Phys. Lett. {\bf B428}, 105 (1998), hep-th/9802109.

\bibitem{Randall}
L. Randall and R. Sundrum,
{\sl An alternative to compactification},
Phys. Rev. Lett. {\bf 83}, 4690 (1999), hep-th/9906064.

\bibitem{wittensusskind}
L. Susskind and E. Witten,
{\sl The Holographic Bound in Anti-de Sitter Space},
hep-th/9805114.

\bibitem{Garriga}
J. Garriga and T. Tanaka,
{\sl Gravity in the Randall-Sundrum Brane World},
Phys. Rev. Lett. {\bf 84}, 2778 (2000), hep-th/9911055.

\bibitem{Muck}
W. M\"uck, K. S. Viswanathan and I. V. Volovich,
{\sl Geodesics and Newton's law in brane backgrounds},
hep-th/0002132.

\bibitem{Capper3}
D. M. Capper and M. J. Duff,
{\sl Trace anomalies in dimensional regularization},
Nuovo Cimento {\bf A23}, 173 (1974).

\bibitem{Deser}
S. Deser. M. J. Duff and C. J. Isham,
{\sl Non-local conformal anomalies},
Nucl. Phys. {\bf B11}, 45 (1976).

\bibitem{Duff3}
M. J. Duff,
{\sl Observations on conformal anomalies},
Nucl. Phys. {\bf B125}, 334 (1977).

\bibitem{Duff4}
M. J. Duff,
{\sl Twenty years of the Weyl anomaly},
Class. Quantum. Grav. {\bf 11}, 1387 (1994), hep-th/9308075.

\bibitem{Henningson}
M. Henningson and K. Skenderis,
{\sl The holographic Weyl anomaly},
JHEP {\bf 9807}, 023 (1998), hep-th/9806087.

\end{references}
\end{document}